\documentclass[conference]{IEEEtran}

\ifCLASSINFOpdf
   \usepackage[pdftex]{graphicx}
\else
   \usepackage[dvips]{graphicx}
\fi
\usepackage[cmex10]{amsmath}

\usepackage{graphics} 
\usepackage{graphicx}
\usepackage{subcaption}
\usepackage{array}
\usepackage{cite}
\usepackage{epsfig} 
\usepackage{amsmath} 
\usepackage{amssymb}  
\usepackage{amsthm}
\usepackage{color}
\usepackage{float}

\usepackage{algorithmic}
\usepackage{url}
\usepackage{calrsfs}
\usepackage{textcomp}
\usepackage{multirow}
\usepackage{booktabs}
\usepackage{bm} 
\usepackage{hyperref}
\usepackage[noabbrev]{cleveref}

\DeclareMathAlphabet{\pazocal}{OMS}{zplm}{m}{n}
\DeclareMathAlphabet{\mathpzc}{OT1}{pzc}{m}{it}
\newcolumntype{P}[1]{>{\centering\arraybackslash}p{#1}}

\usepackage[linesnumbered,ruled,vlined]{algorithm2e}
\SetKwInput{KwInput}{Input}                
\SetKwInput{KwOutput}{Output}              
\SetKwInput{KwData}{Parameters}              
\usepackage[cal = pxtx, scr = dutchcal]{mathalpha}

\begin{document}

\title{State-of-Charge Aware EV Charging}
\author{\IEEEauthorblockN{Yize Chen}
\IEEEauthorblockA{\textit{Lawrence Berkeley National Laboratory} \\
Berkeley, CA, USA \\
yizechen@lbl.gov\vspace{-40pt}}
\and
\IEEEauthorblockN{Baosen Zhang}
\IEEEauthorblockA{\textit{University of Washington} \\
Seattle, WA, USA \\
zhangbao@uw.edu\vspace{-40pt}}}

\maketitle

\begin{abstract}
Recent proliferation in electric vehicles (EVs) are posing profound impacts over the operation of electrical grids. In particular, due to the physical constraints on charging stations' capacity and uncertainty in charging demand, it becomes an emerging challenge to design high performance scheduling algorithms to better serve charging sessions. In this paper, we design a predictive charging controller by actively incorporating each EV's state-of-charge (SOC) information, which has strong effects on the utilization of dispatchable power during peak hours. Simulation results on both synthetic and real-world EV session and charging demand data demonstrate the proposed algorithm's benefits on maximizing charging throughput and achieving higher rate of feasible charging sessions while satisfying battery and station physical constraints at the same time.
\end{abstract}

\section{Introduction}
EV proliferation is just unfolding worldwide. In the United States, a recent report predicts under medium scenario, EV consumption will increase 30-fold from 2020 to 2050~\cite{mai2018electrification}. However, such rapid electrification of transportation sector presents unprecedented challenge to the charging infrastructure. For uncontrolled EV charging system, the unplanned charging sessions may lead to transformer overloading, reduced equipment lifetime, low charging capacity utilization ratio, and even bring distribution grid operational challenges~\cite{lopes2010integration}. Compared to unmanaged charging, recent report found specifically designed smart charging EVs could not only maximize the usage of charging facilities, but also further reduce greenhouse gas emissions by additional $32\%$~\cite{lipman2020open}. By properly controlling the charging rate for individual EV, system operator can achieve the operational goals such as alleviating over-capacity demand during peak hours, fulfilling more charging requests, or participating in demand-side management and other utility services.

In order to realize highly flexible and reliable charging support for EVs, researchers have looked into the scheduling and operation problem of a charging station. \cite{bitar2016deadline} focused on earliest-deadline-first charging scheme, and proved the average-cost optimality of such policy. If the full information regarding each charging session is beforehand revealed to the charging station operator, the offline optimal charging strategy can be derived by solving an optimization problem~\cite{gan2012optimal, ma2011decentralized}. Online formulation and hardware interface were introduced in \cite{lee2018large}. \cite{alizadeh2016optimal, lee2020pricing} designed pricing schemes to either optimize social welfare or minimize system costs.  There are also learning-based algorithms to tackle the problem of charging session uncertainty and associated resource allocation under congestion~\cite{al2020adaptive,chen2018optimal}. For a survey of EV charging algorithms, we refer readers to \cite{mukherjee2014review}.

\begin{figure}[t]
	\centering
	\includegraphics[width=1.0\linewidth]{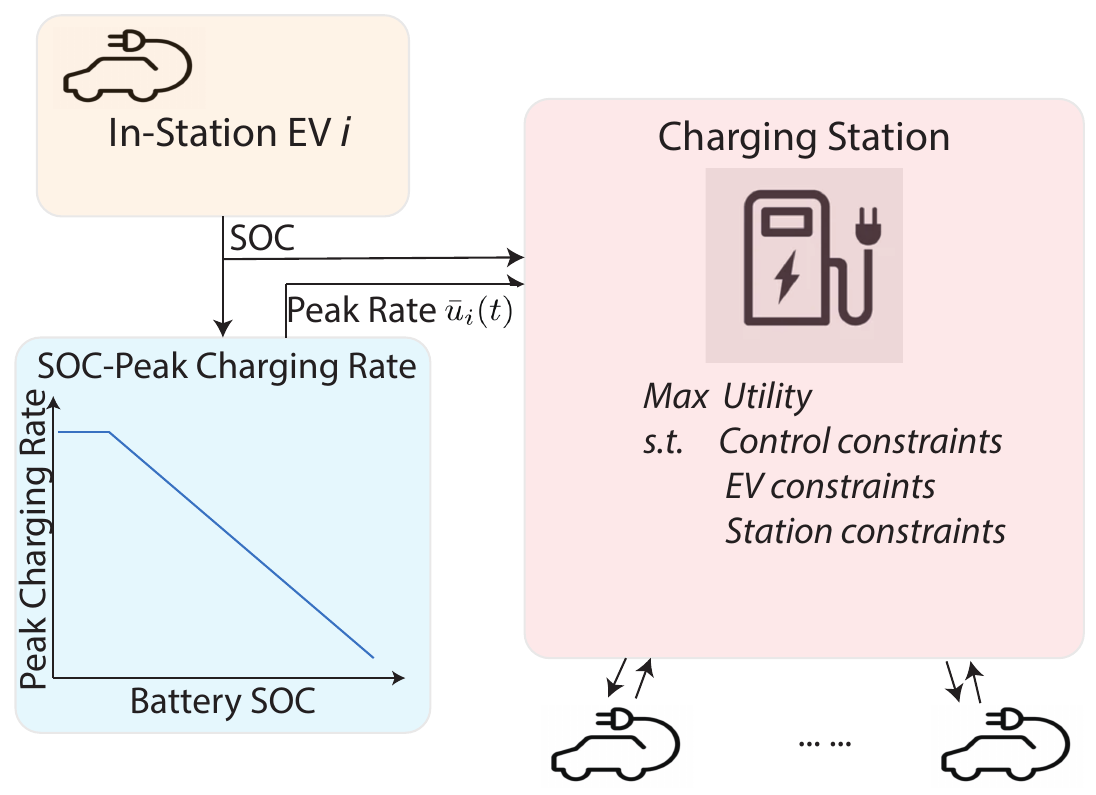}
	\caption{\footnotesize The schematic of proposed algorithm on station-level EV charging control. By explicitly modeling the relationship between peak charging rate ($\bar{u}_i(t)$) and the battery SOC, proposed algorithm is able to better allocate available charging power to individual EVs while avoiding power loss.}
	\label{fig:schematic}
\end{figure}

However, in most of existing charging station models, the physical characteristic of battery peak charging rate with respect to state-of-charge (SOC) is not explicitly taken into consideration. For the typical battery in EV, once the SOC is greater than a relatively small ratio (e.g., $15\%$), the maximum allowed charging power decreases gradually with respect to SOC level due to the electrochemical characteristics~\cite{mukherjee2014review}. Overlooking such peak rate - SOC curve may cause system operator allocating redundant and over-limit power for individual EVs with high SOC. It is thus becoming critical to quantify how battery SOC impact the peak charging rate, which in turn affects the utilization of available power during peak hours.

In this paper, we take the perspective from operating a charging station with unknown future charging demand. We explicitly take EV battery SOC into account, and examine how to incorporate time-variant peak charging rate into charging session scheduling. By focusing on modeling both the public charging stations and EV charging sessions, we propose a fast predictive controller and a tractable solving procedure. Our algorithm is illustrated in Fig. \ref{fig:schematic}. At each timestep, the controller determines the optimal charging rate for each EV in order to fulfill operator's objective such as maximizing total power delivered to the vehicles while satisfying station and battery constraints. The key insight comes from the flexibility provided by the heterogeneous deadlines of charging tasks, while we can successfully defer charging load during peak hours by modeling the variable peak charging rate as a function of battery SOC. In the numerical experiments, we demonstrate that for different setups on peak rate - SOC curve and charging demand, the proposed SOC-aware EV charging scheduler can always accommodate more feasible charging tasks when charging capacity is limited.

\vspace{-10pt}

\section{EV Charging Model}
\label{sec:model}
In this section we describe charging system model, and illustrates how to model both the EV level and station level constraints. Along with the operator's objective, we formulate the optimization problem for scheduling charging sessions, and illustrate the importance of procuring charging flexibility during peak hours.\vspace{-10pt}

\subsection{Charging Station Model}
We consider a single charging station which can provide simultaneous charging services for multiple EVs. Let $\mathcal{V}:=\{1, 2, 3, ...\}$ be the set of all EVs over an optimization horizon $\mathcal{T} := {1, . . . , T }.$ For each EV charging session $i$, it is represented by a 5-element tuple $(x_{i, initial}, x_{i, final}, t_{i, arrival}, t_{i, depart}, \bar{u}_i)$, where $x_{i, initial}$ is the initial state of the battery when arriving at the station, and $x_{i,final}$ is user-specified required SOC before departure. $t_{i, arrival}$ and $t_{i,depart}$ denote the arrival and departure time for vehicle $i$ respectively. For each EV, there is also a hard constraint $\bar{u}_i$ on the EV's peak charging rate, \emph{which is a unique value for each battery and can be time-varying}. We will detail the modeling of such peak charging rate in Section \ref{sec:attack}.

The system operator of the EV charging station is interested in finding a feasible schedule of charging rate $u_i(t), i\in \mathcal{V}, t\in \mathcal{T}$ which can satisfy each customer's charging needs. There are a set of operational constraints which have to be satisfied considering both the charging station and charging sessions physical limits:

\subsubsection{Charging session temporal constraints} For each EV $i$, the charging power is only available when the charging session is live:
\begin{equation}
\label{equ:time}
    u_i(t)=0, \quad t \notin [t_{i,arrival},t_{i, depart}), \; i \in \mathcal{V}.
\end{equation}

\subsubsection{Available charging power} At each timestep, there is a limit on the total power drawn from the charging station:
\begin{equation}
\label{equ:sum_p}
    \sum_{i \in \mathcal{V}}  u_i(t) \leq P(t), \quad t\in \mathcal{T}.
\end{equation}

We note that $P(t)$ can be a limit imposed by the parking garage infrastructure such as transformer capacity or available renewable generations. In this work, we assume there are adequate chargers for onsite EVs.

\subsubsection{Valid charging rate} For each EV $i$, the charging power is nonnegative and always limited by the maximum acceptable charging rate $\bar{u}_i(t)$:
\begin{equation}
\label{equ:u}
    0\leq u_i(t) \leq \bar{u}_i(t), \; i\in \mathcal{V}.
\end{equation}

\subsubsection{EV battery state} For each EV with a live charging session, the battery state is initiated with EV's arrival state, and is updated based upon the current charging rate $u_i(t)$:
\begin{subequations}
\label{equ:state}
\begin{align}
        x_i(0)=&x_{i,initial},\\
    x_{i}(t)=&x_i(t-1)+\delta u_i(t), \; t\geq 1
\end{align}
\end{subequations}
where $\delta$ is a constant with unit of hours of sampling time intervals.

\subsection{Scheduling Objective}
There are several different objectives possibly achieved by the charging facility operator. In this work, we encourage maximizing the sum of each customer's utility throughout $\mathcal{T}$.  Other possible choices include following pre-defined load profile~\cite{lee2018large}, reducing load variance with demand response goals~\cite{gan2012optimal, zhang2016scalable}, or minimizing system cost considering variable price~\cite{alizadeh2016optimal}, while such objectives can be flexibly integrated into our scheduling framework. The utility maximization problem (\texttt{MPC}) can be formulated as follows:
    \begin{subequations}
    \label{equ:MPC}
     \begin{align}
    \max_{\mathbf{u}} \quad & \sum_{t\in \mathcal{T}}\sum_{i\in \mathcal{V}} \log u_i(t) - \lambda \sum_{i\in \mathcal{V}}||x_i(T)-x_{i,depart}||_2^2 \label{equ:obj}\\
    s.t. \quad & \eqref{equ:time}, \eqref{equ:sum_p}, \eqref{equ:u}, \text{and} \eqref{equ:state};
     \end{align}
    \end{subequations}
where $\mathbf{u}$ denotes the collection of $u_i(t), \; i\in \mathcal{V}, \; t \in \mathcal{T}$ and $\lambda$ is a weighting parameter.

The objective \eqref{equ:obj} is a strictly concave function, and it will incur a proportionally fair vector for each timestep, which maximizes the sum of all vechicle's logarithmic utility functions. We choose the objective function to motivate fair sharing of available $P(t)$ and meeting the terminal energy demand at the same time. By solving \eqref{equ:MPC} based on current availability of charging power and connected EVs, we can find optimal charging action which satisfy all operating constraints.

\vspace{-10pt}

\section{SOC Aware Charging Scheduling}
\label{sec:attack}
In this section, we describe how to integrate each vehicle's SOC-dependent peak charging rate into the design of charging sessions scheduling. We would like to emphasize that by leveraging the SOC and peak charging rate information, the proposed algorithm can manage to serve more EVs with higher throughput.

\subsection{Modeling of Peak Charging Rate under Variable SOC}

\begin{figure*}
	\centering
	\includegraphics[width=0.9\linewidth]{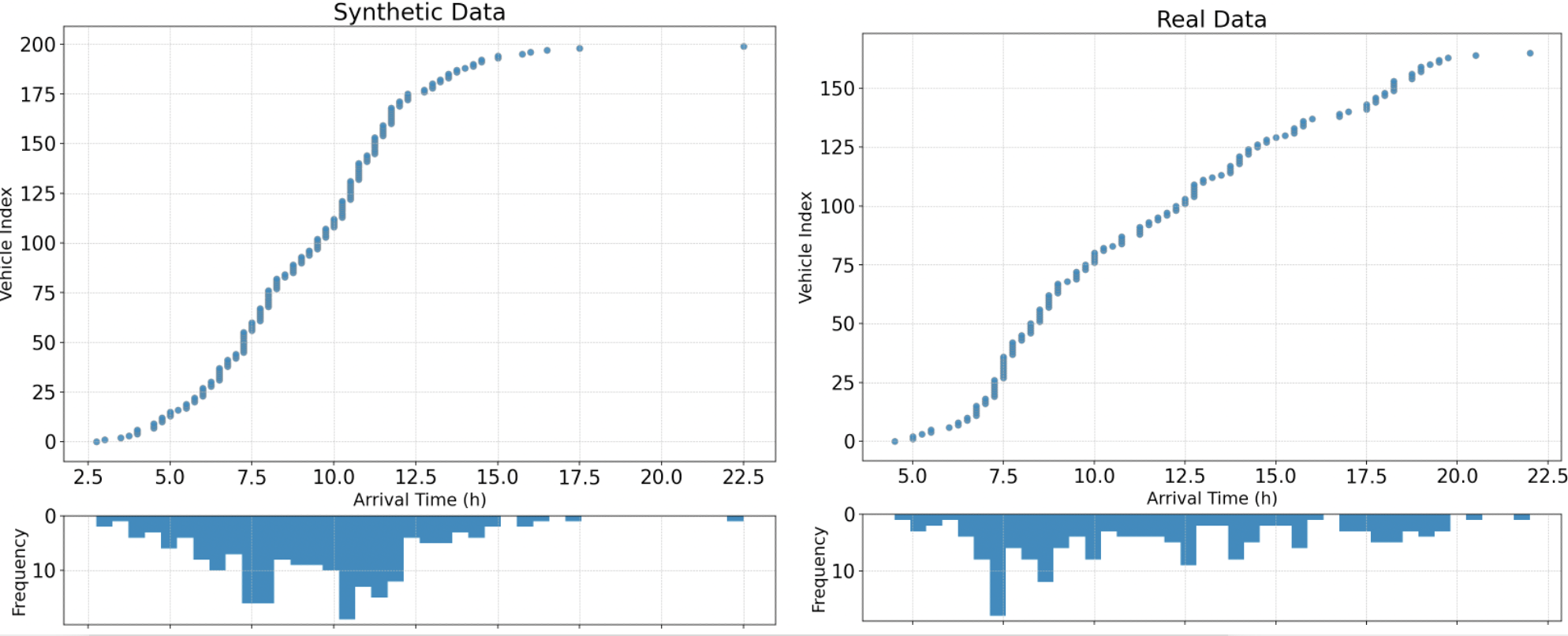}
	\caption{\footnotesize The arrival time statistics for our synthetic data test case and ACN real data. The charging peaks arrive  in the morning.}
	\label{fig:session}
\end{figure*}

One intrinsic limitation of previous work is treating the peak charging rate $\bar{u}_i(t)$ as a time-invariant variable. Under such assumption, solving the ordinary \texttt{MPC} problem \eqref{equ:MPC} leads to inefficient use of charging capacity especially when battery is approaching fully charged states, as the dispatched $u_i(t)$ may exceed the actual peak charging rate. These unutilized energy during congested hours can alternatively meet greater charging demands. To overcome such limitation, we propose to model $\bar{u}_i(t)$'s dependency on SOC using the following function:  

\begin{equation}
\label{equ:bar_u}
    \bar{u}_i(t)=\bar{u}_i^*-\alpha_i \cdot x_i(t-1), \quad i\in \mathcal{V}
\end{equation}
where we term $\alpha_i$ as a ``decaying factor'' to model the peak charging rate decrease with respect to battery SOC. And $\bar{u}_i^*$ is the nominal value for the maximum charging rate. Adding such evolution of $\bar{u}_i(t)$ results to the following optimization problem (\texttt{SOC\_MPC}):

    \begin{subequations}
    \label{equ:SOC_MPC}
     \begin{align}
    \max_{\mathbf{u}} \quad & \sum_{t\in \mathcal{T}}\sum_{i\in \mathcal{V}} \log u_i(t) - \lambda \sum_{i\in \mathcal{V}}||x_i(T)-x_{i,depart}||_2^2 \label{equ:obj_2}\\
    s.t. \quad & \eqref{equ:time}, \eqref{equ:sum_p}, \eqref{equ:u}, \eqref{equ:state}, \text{and} \eqref{equ:bar_u}.
     \end{align}
    \end{subequations}

We also note that such modeling assumption on SOC is practical and can be easily integrated into real-world EV charging stations. Current charging station design realizes the possibility of collecting and utilizing vehicle's SOC information with both hardware and software infrastructure~\cite{BMW}. When an EV plugs in, the vehicle sends information about its current SOC, so that $\bar{u}_i(t)$ can be adjusted in real time for both operator's and EV owner's benefits. 

However, as $\bar{u}_i(t)$ is a function of $x_i(t)$, while $u_i(t)$ is further constrained by $\bar{u}_i(t)$, optimization problem \eqref{equ:SOC_MPC} can not be solved directly using off-the-shelf solvers due to the nested nature of constraints and time-varying bounds in inequality constraints.  In \cite{cao2011optimized}, the authors proposed a heuristic based method to solve the optimization problem involving time-changing $\bar{u}(t)$, yet the optimality and solution feasibility are not guaranteed. We will illustrate our solution procedure to find the exact solution in the next subsection.

\subsection{SOC Aware Adaptive Charging}

For each vehicle $i$, to resolve the dependence between optimization variables and time-varying $\bar{u}_i(t)$, we explicitly write out the relationship between $\bar{u}_i(t)$ and $x_i(t)$ for all timestep:

\[
\begin{bmatrix}
    \bar{u}_{i}(1)\\
        \bar{u}_{i}(2)\\
    \vdots \\
    \bar{u}_{i}(T)
\end{bmatrix} = \bar{u}_i^* \cdot \mathbf{1}-\alpha_i \begin{bmatrix}
   1 &  &  & \\ 
   & 1&  & \\ 
   &  &  \ddots & \\ 
   &  &   & 1 
 \end{bmatrix}\begin{bmatrix}
    x_{i}(0)\\
        x_{i}(1)\\
    \vdots \\
    x_{i}(T-1)
\end{bmatrix}
\]

Then we can further replace \eqref{equ:u} in the vector form $\mathbf{0}\leq \mathbf{u}_i \leq \mathbf{\bar{u}}_i$ where $\mathbf{u}_i, \mathbf{\bar{u}}_i \in \mathbb{R}^T$. The state evolution function \eqref{equ:state} can be also rolling out in the matrix form:

\[
\begin{bmatrix}
    x_{i}(1)\\
        x_{i}(2)\\
    \vdots \\
    x_{i}(T)
\end{bmatrix} = \begin{bmatrix}
    x_{i}(0)\\
        x_{i}(1)\\
    \vdots \\
    x_{i}(T-1)
\end{bmatrix}+
\delta \begin{bmatrix}
    u_{i}(1)\\
        u_{i}(2)\\
    \vdots \\
    u_{i}(T)
\end{bmatrix}
\]

In this way, we can explicitly model all the constraints on EV peak charging rate, EV charging capacity and charging states, and thus we can find the optimal solution for the \texttt{SOC\_MPC} problem with low computation complexity.

\begin{algorithm}[]
\caption{SOC Aware EV Charging}
  \label{algo}
  \KwInput{Charging session information}
  \KwOutput{Charging actions $\mathbf{u}(1),..., \mathbf{u}(T)$}
  \KwData{Charging station capacity $P(t)$}
  \For{t=0,..., T}{
  $\hat{\mathcal{V}_t}:=\{i\in \mathcal{V}|t_{i,arrival}\leq t \; \textbf{AND} \; t_{i,depart}> t \}$; \\
    $\{u_i(t)\}$=\texttt{SOC\_MPC}$(x_{i}(t-1), P(t), t)$\\
    Validate feasibility of $u_i(t)$;\\
    $x_i(t)=x_i(t-1)+\delta u_i (t)$;\\
    Check terminal constraint based on $x_i(t)$ and $x_{i, depart}$;
   }
   Check EV departure and arrival information.
   
\end{algorithm}

The detailed algorithm is listed in Algorithm \ref{algo}. In our implementation, the proposed SOC-aware charging scheme resolves the problem every time when there is a new EV coming/leaving, or the terminal states for required energy are met so that the list of active sessions $\hat{\mathcal{V}}_t:=\{i\in \mathcal{V}|t_{i,arrival}\leq t \; \textbf{AND} \; t_{i,depart}> t \}$ needs to be updated. At time $t$, the predictive controller receive last step's EV states and final required energy $x_{i, depart}$, and implements the action $\mathbf{u}(t)$. We note that the EV's state evolution is always based on actual $\bar{u}_i(t)$, so that over-the-limit charging power will not be used. We will show in the next section such SOC information is valuable especially when there is a congestion in the charging station (e.g., the total available charging power $P(t)$ can not meet all EVs to charge at their full capacity), and can allocate limited power more efficiently.

\vspace{-10pt}

\section{Simulations}
\label{sec:case}
In this section, we will evaluate the performance of proposed SOC-aware EV charging mechanism (\texttt{SOC\_MPC}), and compare it against other algorithms such as equal share (ES)~\cite{kay1988fair},  earliest-deadline-first (EDF)~\cite{stankovic2012deadline, bitar2016deadline}, and model predictive control (MPC)~\cite{lee2018large} (without considering SOC information). We make the code for EV station modeling and proposed algorithm publicly available\footnote{\href{https://github.com/chennnnnyize/State-Demand_Aware_EV_Charging}{github.com/chennnnnyize/State-Demand\_Aware\_EV\_Charging}}.

\begin{figure}[t]
	\centering
	\includegraphics[width=1.0\linewidth]{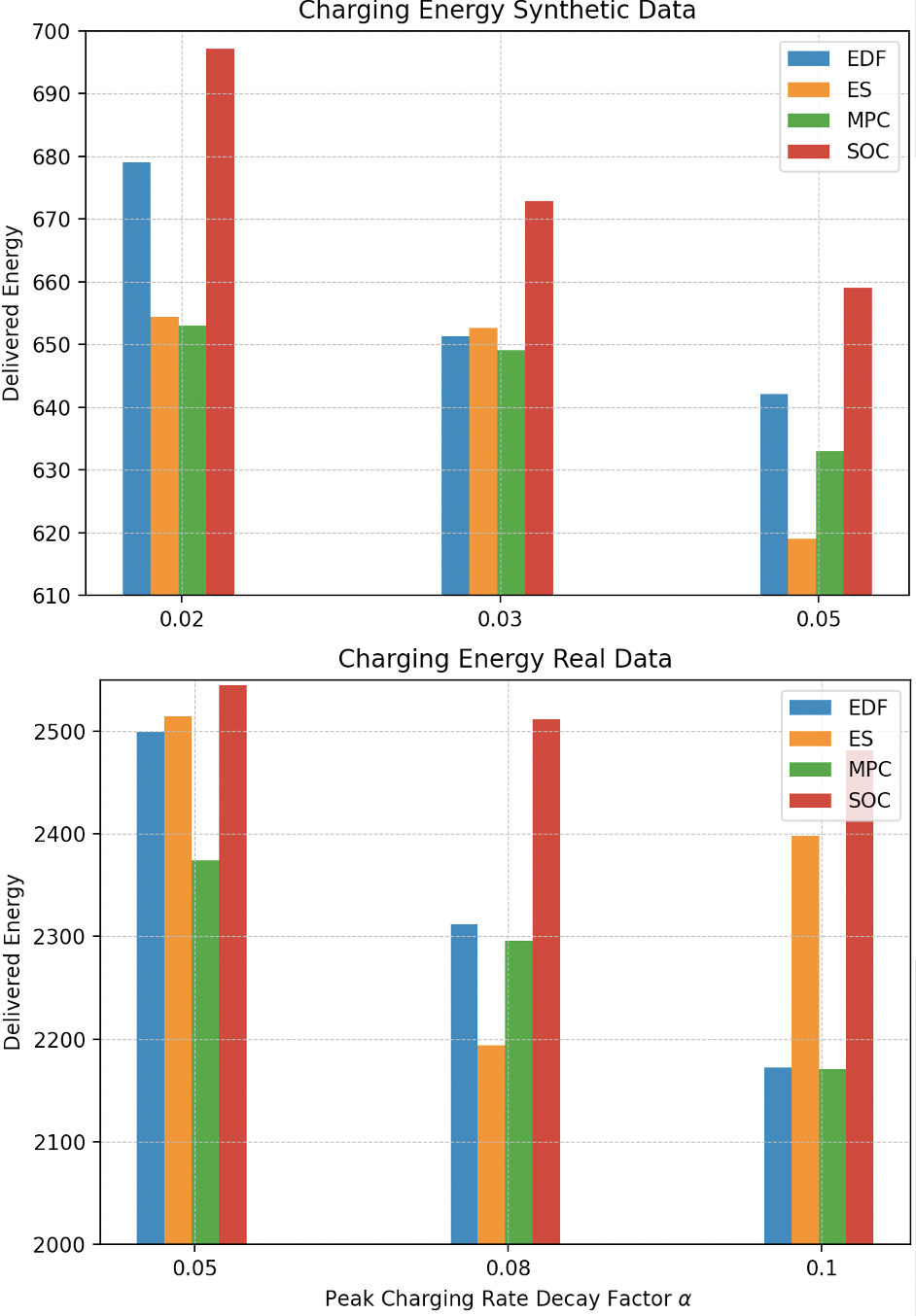}
	\caption{\footnotesize Comparison on the total delivered energy of four methods on two datasets and varying battery $\alpha$ setup. \texttt{SOC\_MPC} can maximize the overall throughput for all test cases.}
	\label{fig:schematic}
\end{figure}
\vspace{-10pt}
\subsection{Test Cases and Simulation Setup}
We use both synthetic dataset and real-world EV charging station data from ACN-Data~\cite{lee2019acn}. In the synthetic data, we model the commuting pattern for a workspace charging station using a Poisson process, and randomly sample the initial SOC and required energy for each individual EV. For the ACN-Data, we process the EV data in the week of September 5th to September 11th in 2021 for the Caltech site. The charging session temporal distribution is illustrated in Fig. \ref{fig:session}. For the synthetic case, we simulate charging scheduling with $\alpha=\{0.02, 0.03, 0.05\}$ and for the real data case, we have $\alpha=\{0.05, 0.08, 0.1\}$. We choose charging station power capacity $P(t)$ such that offline feasible solution can be found given the EV charging demand. We notice the synthetic data has a sharper morning peak with regard to charging sessions.

We include a brief introduction of other implemented EV charging schemes for method comparisons:
\begin{itemize}
\item The ES algorithm allocates available charging power supply equally to all active EV charging sessions at each timestep while satisfying each EV's minimum and maximum charging rate. 
\item The EDF algorithm assigns $\bar{u}_i$ to EVs with earliest deadlines while keeping the sum of charging power less or equal to $P(t)$.
\item The ordinary  \texttt{MPC} algorithm iteratively solves \eqref{equ:MPC} with a fixed $\bar{u}_i$, and other settings are kept the same as \texttt{SOC\_MPC}.
\end{itemize}

For all tested cases and implemented algorithms, we update the charging status when either an EV connects, departs or the charging session terminal constraints have been met. Without loss of generality, we choose simulation interval as $15$ minutes, and set $\delta=1$ in all our simulation cases. We look into three different settings on the peak charging rate-SOC parameter $\alpha$\footnote{The choice of $\alpha$ is based on the battery capacity, and thus we choose different set of values for the two datasets.}, and illustrate the value of including $SOC$ information into efficient EV charging algorithm design. We use CVXPY~\cite{diamond2016cvxpy} to solve the optimization problem and find MPC solutions.

\begin{figure}[t]
	\centering
	\includegraphics[width=1.0\linewidth]{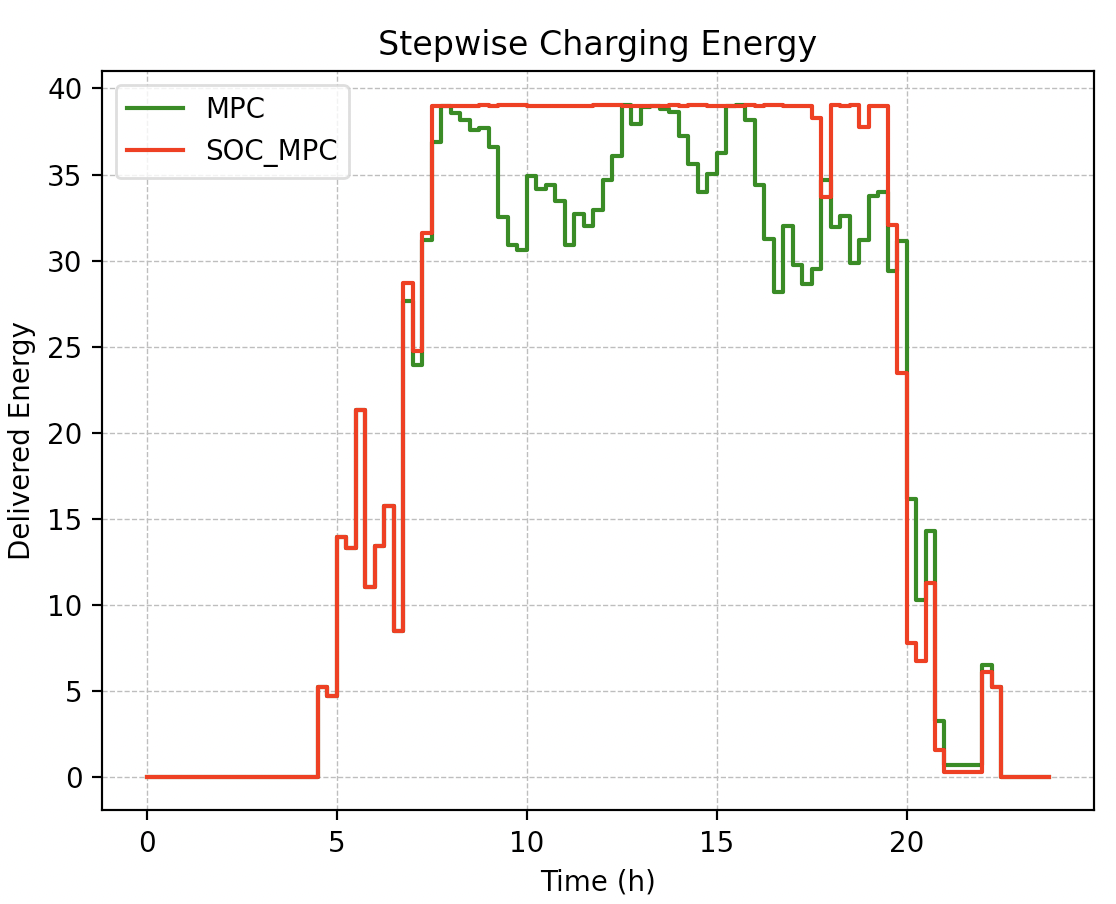}
	\caption{\footnotesize The comparison of stepwise delivered energy between \texttt{MPC} and \texttt{SOC\_MPC} on the ACN dataset with $\alpha=0.1$.\vspace{-10pt}}
	\label{fig:step}
\end{figure}

\vspace{-10pt}
\subsection{Performance Evaluation}
We first compare the amount of delivered energy using all four algorithms. In all settings with different $\alpha$ for both datasets, we find proposed algorithm can always deliver the most energy to EVs, and the average performance gain with respect to ordinary \texttt{MPC} is $10.27\%$. ES and EDF algorithms are not performing stably when faced with different settings of EV arrival process and peak charging rate decay. The EDF algorithm are performing well when the charging demand is low, such that charging sessions with earliest deadline can be always satisfied. Once the charging demand are high while EDF is dispatching too much power based on schedule priority, a portion of the power is not fully utilized on the EVs with high SOC. The ES algorithm is performing more evenly under different setting of $\alpha$, but can not produce many feasible charging sessions with regard to the final energy requests $x_{i, depart}$, since many sessions requesting higher quantity of energy during peak hours are only allocated small amount of power. This illustrates that when there are limited charging power capacity, EDF and ES are not ideal policies in terms of maximizing the usage of charging power or satisfying individual EV charging needs.

Performance gain achieved by \texttt{SOC\_MPC} majorly comes from better utilization of total available power $P(t)$ during peak hours. The awareness of SOC lets the algorithm distributes the charging power to more EVs rather than allocating large amount of power to single EV. This is further illustrated in Fig. \ref{fig:step}, where during peak hours with multiple simultaneous charging sessions, \texttt{SOC\_MPC} is able to make use of almost all of the available charging power.  On the other hand, \texttt{MPC} finds charging actions that are over the charging limit $\bar{u}_i(t)$. Similar deficiencies are observed for ES and EDF algorithm, as EDF and ES often allocate more-than-allowed power to batteries which are almost fully charged.

\begin{figure}[t]
	\centering
	\includegraphics[width=1.0\linewidth]{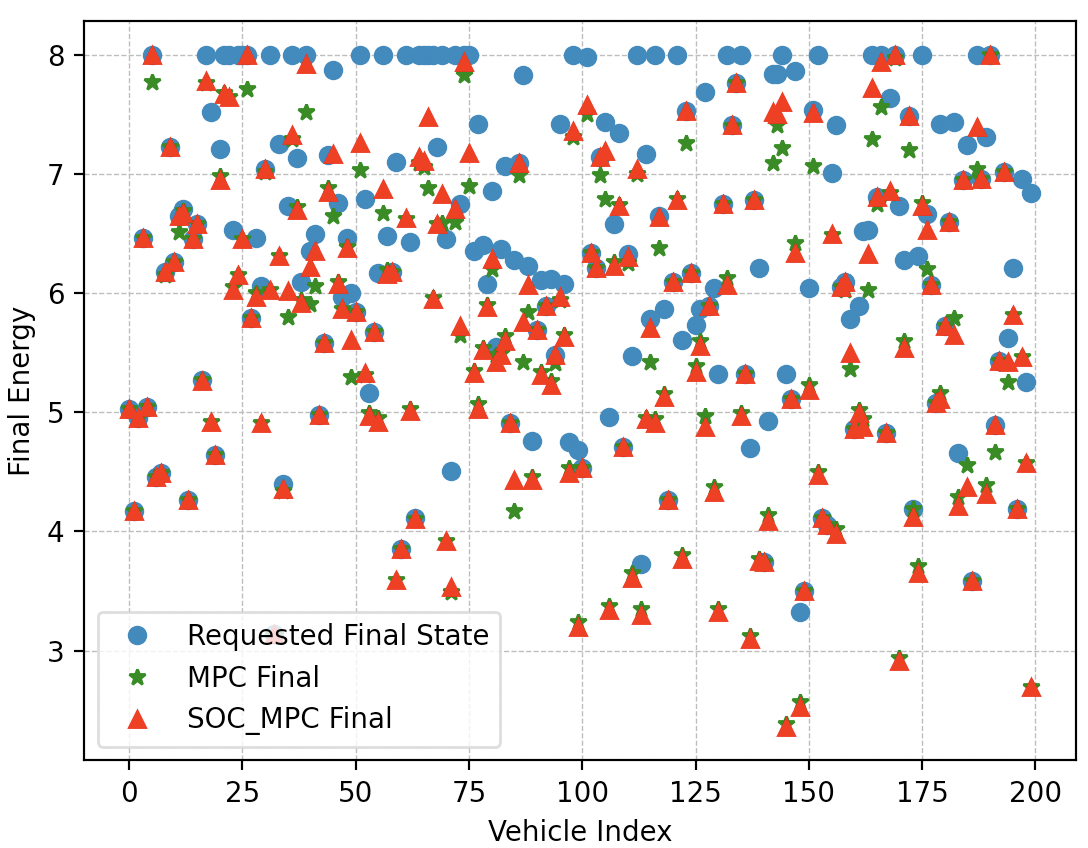}
	\caption{\footnotesize The final battery states on the synthetic dataset with $\alpha=0.05$.}
	\label{fig:states}
\end{figure}

We also compare the final states of batteries in Fig. \ref{fig:states} between the \texttt{MPC} and \texttt{SOC\_MPC} algorithm. We can observe that for almost all vehicles, the final battery states by using the proposed algorithm are higher than using standard \texttt{MPC} controller. This also shows the promise that SOC information can help design algorithm that brings benefits for EV customers.

\section{Conclusion and Discussion}
\label{sec:conclusion}
In this work, we show that state-of-charge information for EV batteries  are valuable for designing charging station algorithms in terms of maximizing energy usage and meeting more charging session requests. By explicitly considering the EV peak charging rates with respect to SOC as time-varying constraints, we design an EV charging controller to better allocate limited charging power particularly during peak hours. This calls for a more accurate modeling of the battery characteristics especially the peak charging rate given different SOC values.  In the future work, we will investigate the predictive model for the future uncertain charging demand, and also analyze how to harness flexibility as grid resources via smart charging interfaces.

\bibliographystyle{IEEEtran}
\bibliography{bib}

\end{document}